\documentclass[twocolumn,aps,prl,reprint,groupedaddress]{revtex4-1} 
\usepackage{graphicx}
\usepackage{amsmath,amssymb}
\usepackage{textcomp,siunitx}
\usepackage{color}

\newcommand{\ket}[1]{\left\lvert #1 \right\rangle}%
\newcommand{\avg}[1]{\left\langle #1 \right\rangle}%
\renewcommand{\vec}[1]{\boldsymbol{#1}}

\usepackage{xpatch}

\xpatchcmd{\thebibliography}{%
  \usecounter{enumiv}%
}{%
  \usecounter{enumiv}%
  \setcounter{enumiv}{\value{20}}%
}{}{}

\begin{document}

\title{Probing quench dynamics across a quantum phase transition into a 2D Ising antiferromagnet}
\author{Elmer Guardado-Sanchez, Peter T. Brown, Debayan Mitra, Trithep Devakul, David A. Huse, Peter Schau\ss, Waseem S. Bakr}
\email[]{wbakr@princeton.edu}
\affiliation{Department of Physics, Princeton University, Princeton, New Jersey 08544, USA}

\date{\today}

\begin{abstract}
Simulating the real-time evolution of quantum spin systems far out of equilibrium poses a major theoretical challenge, especially in more than one dimension. We experimentally explore the dynamics of a two-dimensional Ising spin system with transverse and longitudinal fields as we quench it across a quantum phase transition from a paramagnet to an antiferromagnet. We realize the system with a near unit-occupancy atomic array of over 200 atoms obtained by loading a spin-polarized band insulator of fermionic lithium into an optical lattice and induce short-range interactions by direct excitation to a low-lying Rydberg state. Using site-resolved microscopy, we probe the correlations in the system after a sudden quench from the paramagnetic state and compare our measurements to exact calculations in the regime where it is possible. We achieve many-body states with longer-range antiferromagnetic correlations by implementing a near-adiabatic quench and study the buildup of correlations as we cross the quantum phase transition at different rates.
\end{abstract}

\maketitle


Lattice quantum spin models serve as a paradigm for exploring a range of many-body phenomena including quantum phase transitions \cite{Sachdev1999,Tanaka2017}, equilibration and thermalization \cite{Blass2016,Mondaini2016}, and quench dynamics \cite{Eisert2015,Sengupta2004,Barmettler2009,Rieger2011,Calabrese2011,Abeling2016}. While there exists a variety of well-developed theoretical techniques to study the equilibrium properties of quantum spin systems \cite{Avella2013,Linden1992,White1992,Schollwoeck2005,Hallberg2006,Tang2013,Orus2014}, the toolkit for simulating real-time dynamics of these systems is rather limited and can only capture the evolution for short times, especially for systems in more than one dimension \cite{Avella2013,Vidal2004,White2004,Hazzard2014}. Recent advances in the field of quantum simulation have introduced several experimental platforms where the dynamics of quantum spin systems can be measured over long evolution times, providing much needed benchmarks for testing uncontrolled theoretical approximations. Examples of such platforms include trapped ions \cite{Kim2010,Britton2012,Bohnet2016}, polar molecules \cite{Yan2013}, Rydberg atoms \cite{Schauss2012,Maller2015,Jau2016,Labuhn2016,Bernien2017}, magnetic atoms \cite{DePaz2013,Baier2016} and atoms interacting through superexchange in optical lattices \cite{Fukuhara2013a,Boll2016,Cheuk2016,Parsons2016,Drewes2017,Brown2017}.


In this work, we explore the dynamics of a two dimensional quantum Ising model using a nearly defect-free array of neutral atoms which are coupled with laser light to a low-lying Rydberg state in an optical lattice \cite{Gallagher1994}. The spin coupling in the model arises due to a van der Waals interaction between atoms in the Rydberg state. If one atom is in a Rydberg state, the excitation of another atom to a Rydberg state is strongly suppressed within a blockade radius $R_b$ \cite{Singer2004,Tong2004,Heidemann2008a,Urban2009,Gaetan2009}. This is because the interaction between the Rydberg atoms within this radius is much larger than the laser coupling strength. Previous experiments in 2D arrays have studied the regime $R_b \gg a_{l}$, where $a_{l}$ is the lattice spacing \cite{Schauss2012, Labuhn2016}. In this regime, the Rydberg blockade makes it difficult to access many-body states with a large Rydberg fraction. This significantly reduces the size of the relevant Hilbert space of the system from the maximum possible size of $2^N$, where $N$ is the number of sites, rendering simulation of the quantum dynamics feasible for the experimentally realized system sizes. Here we focus on the regime $R_b \sim a_{l}$, where there is no such reduction of the Hilbert space size. This regime allows us to study quench dynamics across a quantum phase transition between a paramagnet and an antiferromagnet with broken $\textrm{Z}_2$ symmetry. Recently, Rydberg atoms in rearrangeable optical tweezers have explored this regime in 1D chains \cite{Bernien2017} and rings \cite{Labuhn2016}.

\begin{figure}
    \includegraphics[width=3in]{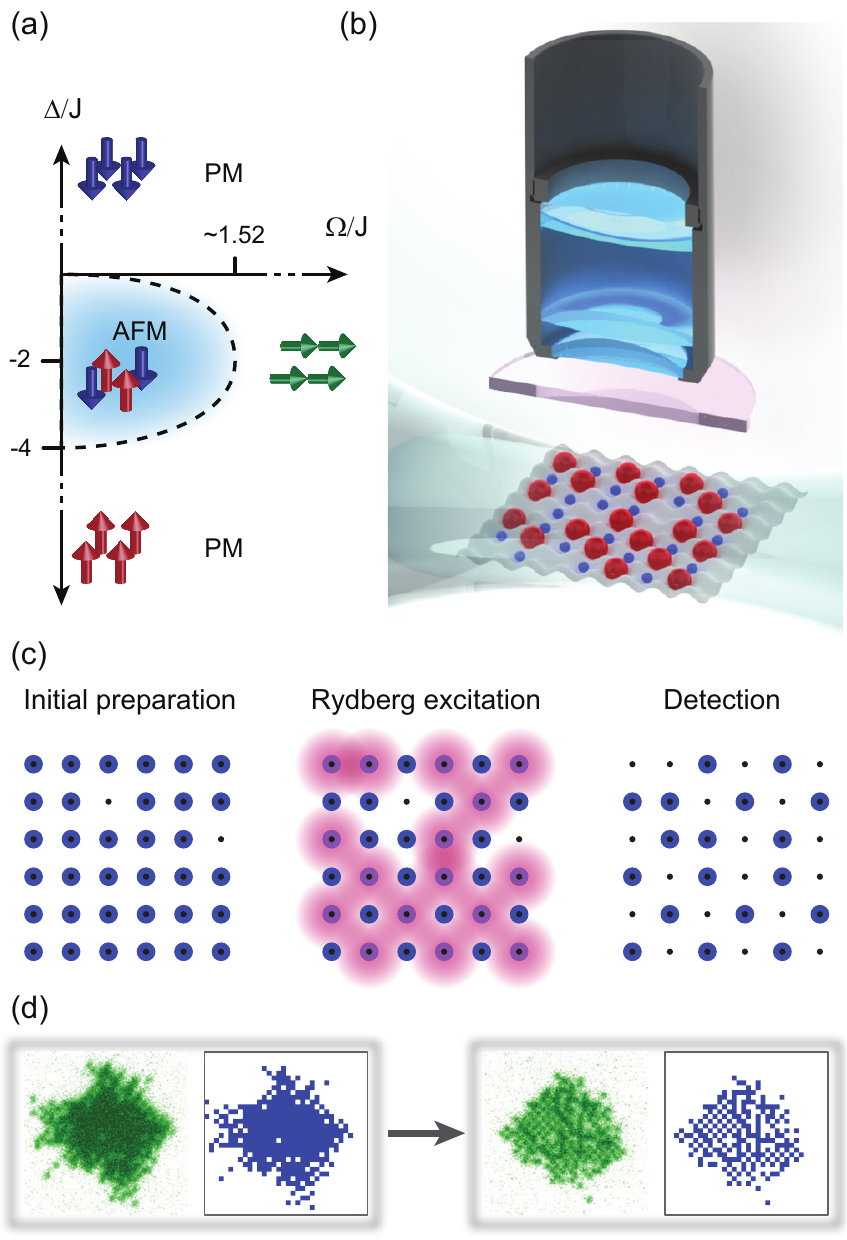}
    \caption{{\bfseries Realization of a 2D quantum Ising model with Rydberg atoms in an optical lattice.} (a) Phase diagram of a 2D quantum Ising model with nearest neighbor coupling $J$. This is an approximate phase diagram of our Rydberg system when the blockade radius is comparable to the lattice spacing. Transverse and longitudinal fields are controlled by the Rabi frequency $\Omega$ and laser detuning $\Delta$, respectively. There is only one ordered  phase, the antiferromagnet (AFM). Outside of this region there is a paramagnetic (PM) phase where the spins align with the field. (b) Experimental setup consisting of a 2D array of atoms at the focus of a high-resolution objective, capable of resolving individual sites of the lattice. Atoms in the ground state (small blue spheres) are directly coupled to the 23$P$ Rydberg state (large red spheres) with 230 nm laser light. (c) Typical atom configurations at different stages of the experiment. The initial state consists of an array of atoms in the electronic ground state (blue, left). This state is quenched into an antiferromagnetic state (Rydberg atoms in red, center). By increasing the lattice depth, Rydberg atoms are lost and only the ground state atoms are imaged (right). (d) Raw fluorescence images of an initial (left) and a post-quench (right) configuration with strong antiferromagnetic correlations, together with reconstructed images (each blue pixels depicts a detected atom in the ground state).\label{fig:fig1}}
\end{figure}

We realize a quantum Ising spin system with an array of $^6$Li atoms in an optical lattice with near unit-occupancy. The lattice is deep enough to suppress tunneling over the timescale of the experiments.  We prepare all the atoms in the same hyperfine ground state $\left|\downarrow\right\rangle$. Interactions are introduced by globally coupling the atoms with a single laser field to a Rydberg state $\left|\uparrow\right\rangle$. The van der Waals interaction between atoms in the Rydberg state is isotropic and takes the form $V_{ij} = C_6/|\textbf{r}_i - \textbf{r}_j|^6$. The Hamiltonian of the system is given by
 \begin{equation} 
H = \Omega
  \sum_{\vec i} \hat S^x_i + \sum_i (\mathcal{I}_i- 
  \Delta) \hat S^z_i + \sum_{i \neq j} \frac{V_{ij}}{2} \hat S^z_i\hat S^z_j\; . 
\end{equation} \label{eq:rydberg}
Here $S^\alpha_i$ are the spin $1/2$ operators for the $i$th lattice site and $\alpha = x,y,z$. The first two terms of this Hamiltonian describe transverse and longitudinal magnetic fields that couple to the pseudospin. The Rabi frequency $\Omega$  that drives a transition between the ground and the Rydberg state for an isolated atom determines the transverse field, while the detuning $\Delta$ of the laser frequency from atomic resonance determines the longitudinal field.  $\mathcal{I}_i = \sum_{j,(i \neq j)} \frac{V_{ij}}{2}$ can be taken as a site independent detuning in a large system as ours. We work with an attractively interacting ($V_{ij} < 0$) Rydberg state \cite{SuppOnline}. In the absence of the fields, the Hamiltonian's most excited state is a classical antiferromagnet, which is the ground state of the Hamiltonian $\tilde{H}=-H$. For $R_b = (C_6/\Omega)^{1/6}\gg a_{l}$, the ground state phase diagram of $\tilde{H}$ in $\Omega/\Delta$ parameter space contains multiple Rydberg crystalline phases with different Rydberg atom fractions \cite{Fendley2004,Pohl2010a,Sela2011,Bernien2017}. However for $R_b \sim a_{l}$, the regime we study in this experiment, $\tilde{H}$ can be approximated by a nearest-neighbor Ising Hamiltonian with coupling $J=C_6/a_l^6$. A phase diagram for this model is shown in Fig.~\ref{fig:fig1}(a) and has only one ordered phase, the antiferromagnet \cite{Yanase1976,Hamer1984}. The initial state in the experiment is the paramagnetic ground state of $\tilde{H}$ for positive detuning $\Delta \gg J \gg \Omega$. In this work, we quench the system from this initial state to the antiferromagnet with varying degrees of adiabaticity and study the ensuing dynamics of the spin correlations.

We prepare nearly defect-free 2D arrays of atoms by taking advantage of Pauli blocking in a highly spin-imbalanced degenerate Fermi gas loaded into a square optical lattice (for details see \cite{Brown2017}). The spin mixture consists of the first $\ket{1}$ and third $\ket{3}$ lowest hyperfine ground states of $^6$Li, with $\ket{1}$ as the majority. The minority atoms, needed to thermalize the gas while loading into the lattice, are subsequently removed with a pulse of resonant light. We focus our analysis on an annular region with outer (inner) radius of 9 (4) sites where the average occupancy of the remaining state $\ket{1}$ atoms, measured from repeated preparations of the system, is maximal and corresponds to \SI{95.7(4)}{\%}.

We couple the state $\ket{1}$ atoms to the 23$P$ ($m_l=0,m_s=-1/2,m_I=1$) Rydberg state using single-photon excitation with an ultraviolet (UV) laser at \SI{230}{\nm}. The experiments are performed at a bias magnetic field of 595 G  pointing orthogonal to the 2D layer, allowing us to address a single $\ket{m_l,m_s,m_I}$ Rydberg state. Up to \SI{60}{\mW} of UV light is available from a frequency-quadrupled diode-laser system. The light is $\pi$-polarized and focused to a waist of $70~\mu$m. The intensity and the frequency of the light can be changed rapidly to control the time dependence of the transverse and longitudinal fields in the Hamiltonian \cite{SuppOnline}. 

The atoms are located at the focus of a high resolution objective that can resolve individual sites of the optical lattice (Fig.~\ref{fig:fig1}(b)). The Rydberg dynamics take place in a lattice of depth 55$E_R$, where $E_R = (\pi \hbar)^2/2ma^2_{l}$ is the recoil energy and $a_{l} = 1064$ nm$/\sqrt{2}$. We image the distribution of ground state atoms after removing Rydberg atoms with an efficiency of \SI{90(3)}{\%} by increasing the lattice depth to $2500 E_R$, leading to rapid photoionization or expulsion of the anti-trapped Rydberg atoms (Fig.~\ref{fig:fig1}(c)). We obtain site-resolved fluorescence images of the ground state atoms by collecting $\sim1000$ photons per atom scattered from laser beams in a Raman cooling configuration~\cite{Brown2017}.

We calibrate the transverse and longitudinal fields of the Hamiltonian using sparse clouds where the average spacing between atoms is much larger than $R_b$. The location of the Rydberg resonance ($\Delta=0$) is determined by finding the laser frequency which maximizes atom loss during a long exposure to the UV light, since atoms in the Rydberg state experience an anti-trapping optical potential. The Rabi frequency $\Omega$ is determined by measuring single atom Rabi oscillations, and we attain a maximum Rabi frequency $\Omega_{max}=h\times\SI{5.4(1)}{\MHz}$ \cite{SuppOnline}. $\Omega$ varies $\SI{4.9(3)}{\%}$ over the region of interest due to the Gaussian intensity profile of the UV beam. The $C_6$ coefficient, which determines the strength of the van der Waals interaction, depends strongly on the principal quantum number. We obtain a theoretical $C_6/h = \SI{-1.92(6)}{\MHz\,\um^6} = \SI{-10.6(3)}{\MHz}~a_l^6$ for the $23P,m_l=0$ state at an offset field of \SI{595}{G} ~\cite{SuppOnline}. The angular dependence of the interaction potential in the $P$-state is unimportant in our experiments since the magnetic quantization axis is orthogonal to the plane of the lattice, leading to an isotropic interaction for atoms in the 2D plane. For these parameters, $\Omega_{max}, J \gg h/\tau$, where $\tau\sim20~\mu$s is the lifetime of the Rydberg state \cite{Beterov2009}, leading to negligible decay over the relevant timescales.

\begin{center}
\begin{figure*}[h]
    \includegraphics[width=5.3in]{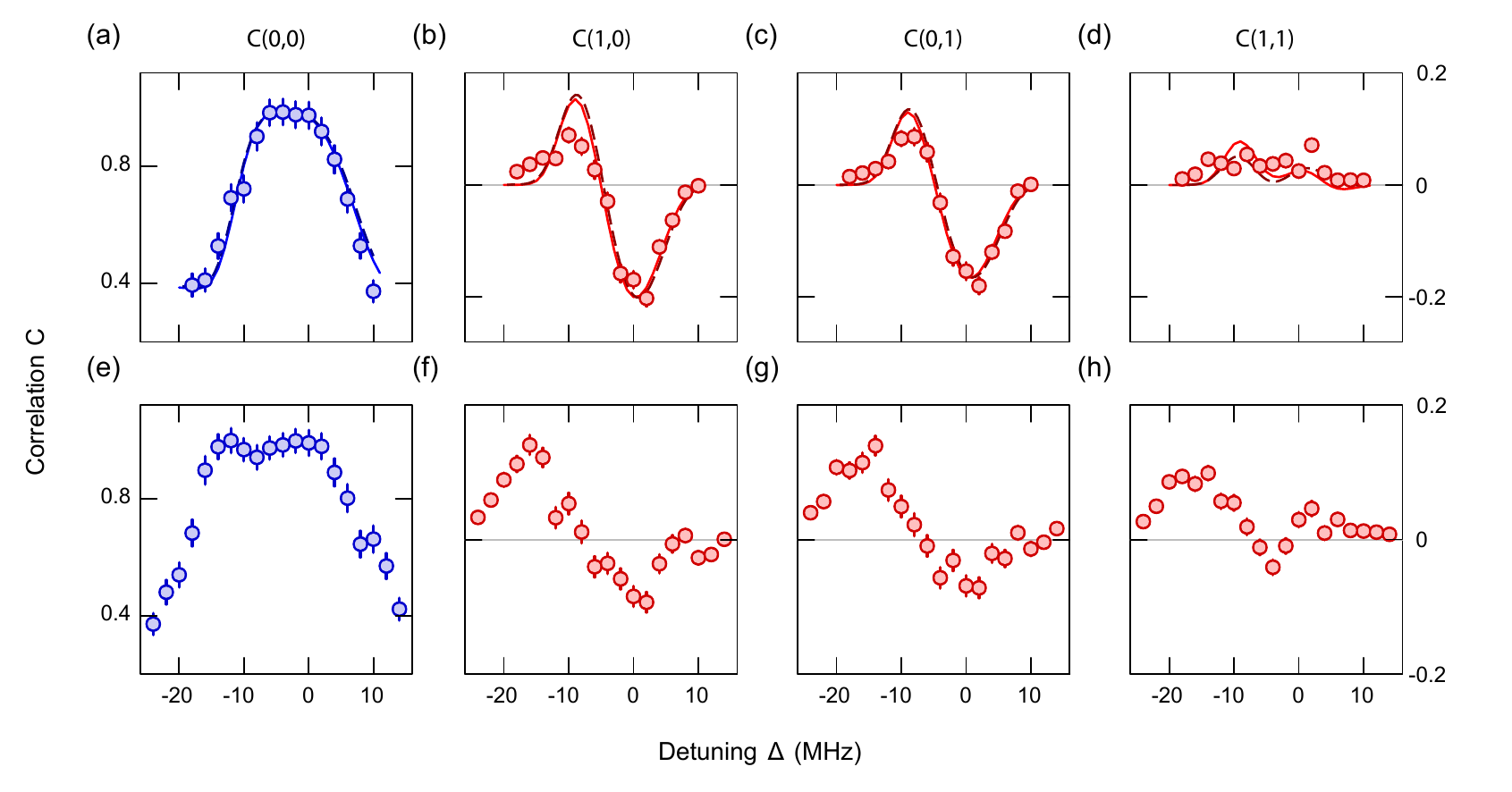}
    \caption{{\bfseries Sudden quench dynamics.} (a-d) Spin correlations after a sudden quench with $\Omega T/h = \pi/2$ ($\Omega = h\times\SI{4.05(2)}{MHz}$) at various detunings $\Delta$. The correlators shown are $C(0,0)$ (a), $C(1,0)$ (b), $C(0,1)$ (c), and $C(1,1)$ (d). For comparison we show the fits to dynamics computed with NLCE (solid line) and exact diagonalization on a $4\times4$ lattice with open boundary conditions (dashed line). (e-h) Spin correlations after a longer quench of $\Omega T/h = 3\pi/2$ ($\Omega =  h\times\SI{5.3(1)}{\MHz}$) at various detunings. \label{fig:fig2}}
\end{figure*}
\end{center}

We first study dynamics in the Ising system after a sudden quench, where the transverse field is switched on quickly compared to $h/\Omega$. The system is initially in a product state, with all spins in $\ket{\downarrow}$, and we image the atoms after an evolution time $T$. From the images, we extract the spin correlators $C(\mathbf{r}) =  4\avg{ S^z_{\mathbf{i}} S^z_{\mathbf{i}+\mathbf{r}} }_c = 4(\avg{ S^z_{\mathbf{i}} S^z_{\mathbf{i}+\mathbf{r}} } - \avg{ S^z_{\mathbf{i}} }\avg{S^z_{\mathbf{i}+\mathbf{r}} })$. The correlators $C(0,0),C(1,0),C(0,1)$ and $C(1,1)$ are shown in Fig.~\ref{fig:fig2}(a)-(d) for $\Omega T/h = \pi/2$ ($\Omega = h\times\SI{4.05(2)}{\MHz}$) and varying detuning $\Delta$. The correlator $C(0,0)$ is linked to the magnetization as $C(0,0)=1 - 4\avg{S^z_{\mathbf{i}}}^2$. We observe a change in the sign of the nearest neighbor correlations as the detuning $\Delta$ is varied. 

For such short times, the correlations remain short-range and the dynamics can be calculated. We implement a dynamical version \cite{SuppOnline,White2017} of the numerical linked cluster expansion (NLCE) to compare with our results \cite{Rigol2007,Tang2013}. The dynamics is computed on clusters of increasing size (the ``order" of the expansion) and the results are expected to converge if the correlation length is smaller than the cluster size. The 11th order NLCE results for the on-site and nearest-neighbor correlations are fit to the measured correlations after the quench with two free parameters: the van der Waals interaction coefficient $C_6$ and a scaling factor $\alpha$ corresponding to the Rydberg imaging efficiency. The NLCE dynamics calculations take into account interactions up to next-nearest neighbors and experimental imperfections including the finite rise and fall time of $\Omega$ and \SI{2.8}{\%} anisotropy of the lattice spacing \cite{Brown2017}, which translates to an \SI{18}{\%} anisotropy of the interactions on the nearest neighbor sites. We also compare the data to exact diagonalization results on a $4\times4$ lattice. From these fits, we obtain an experimental $C_6/h = \SI{-1.1(1)}{\MHz\,\um^6} = \SI{-6.0(3)}{\MHz}~a_l^6$ and a scaling factor $\alpha = 0.89(1)$, which agrees with the expected detection efficiency \cite{SuppOnline}. The fitted value of $C_6$ is about $40\%$ lower than the theoretically calculated $C_6$, which has possible systematic errors due to uncertainties in the matrix elements in lithium, in particular at high magnetic fields, and finite wavefunction size of the atoms on the lattice sites \cite{SuppOnline}.

To go beyond the regime where the dynamics can be calculated, we perform a longer quench with $\Omega T/h = 3\pi/2$. The extracted correlators are shown in Fig. \ref{fig:fig2}(e)-(h). In this case, even the next-nearest neighbor correlations exhibit a zero crossing as a function of detuning, showing that the system is building up longer range correlations. The different NLCE orders already stop converging at much earlier times. Thus, this data presents a challenge to state-of-the-art numerical methods for calculating dynamics.

\onecolumngrid
\twocolumngrid

\begin{figure}
    \includegraphics[width=\columnwidth]{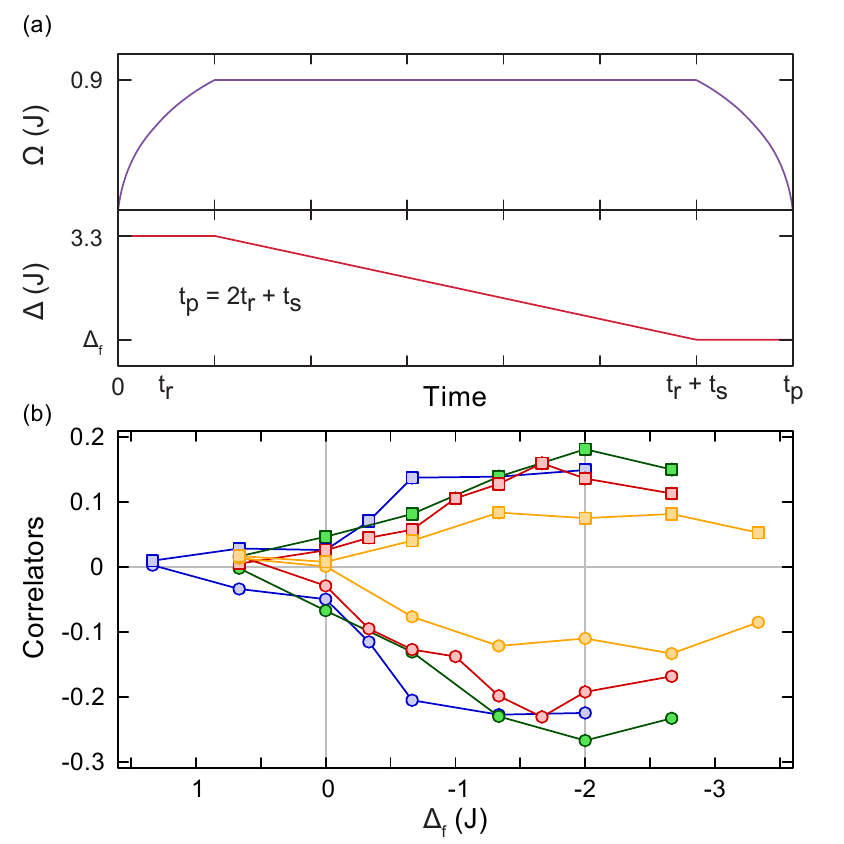}
    \caption{{\bfseries Time evolution of spin correlations during a slow quench across the quantum phase transition.} (a) Time dependence of the Rabi frequency $\Omega$ and detuning $\Delta$ used for the slow quenches. The time for switching on and off the laser coupling $t_r$ was fixed to $0.6h/J$ for all quenches. (b) Buildup of the nearest neighbor (circles) and next-nearest neighbor (squares) correlations during quenches of different speeds $\dot{\Delta} =  8.9J^2/h$ (blue), 4.4$J^2/h$ (green), 2.2$J^2/h$ (red), and 1.6$J^2/h$ (yellow). \label{fig:fig3}}
\end{figure}

To prepare many-body states with longer antiferromagnetic correlations, we investigate a more adiabatic quench scheme \cite{Pohl2010a,Schachenmayer2010,VanBijnen2011}, illustrated in Fig.~\ref{fig:fig3}(a). We start from the same initial state but use a soft switch on and off of the Rabi frequency and a linear ramp of the detuning from $\Delta_i=3.3J$ to a varying $\Delta_f$. During the detuning ramp, the Rabi frequency is fixed at $\Omega_0=0.9(1)J$.  We explore a variety of detuning ramp rates $\dot{\Delta}$ ranging from $1.6J^2/h$ to $8.9J^2/h$. For each $\dot{\Delta}$ we measure correlations at different times in the ramp. Fig.~\ref{fig:fig3}(b) shows the buildup of nearest neighbor and next-nearest neighbor antiferromagnetic correlations as the quantum phase transition is crossed with different speeds. The buildup of antiferromagnetic correlations starts approximately at the time the detuning ramp crosses $\Delta = 0$. For all quench speeds studied, we observe that the correlations reach a maximum at $\Delta/J \sim -2$, as expected from the phase diagram in Fig.~\ref{fig:fig1}(a). The peak value of the correlations initially increases as the quench speed is reduced as one might expect, but then decreases for slower ramps. This is likely due to decoherence playing a more important role in the slower quenches. We studied single spin coherence in our system with a Ramsey sequence in a sparse cloud and did not observe any significant decoherence over the relevant timescales \cite{SuppOnline}. However, strong attractive interactions between atoms in the Rydberg state can lead to many-body decoherence effects resulting from atomic motion. These effects are particularly strong in our system compared to previous optical tweezer experiments \cite{Labuhn2016,Bernien2017} due to the light mass of lithium and the relatively small lattice spacing. Displacement of the atoms, estimated in \cite{SuppOnline}, leads to intrinsic decoherence by changing the spin coupling $J$ and to lesser extent the detuning $\Delta$ because of differential Stark shifts between the ground and Rydberg states induced by the optical lattice potential.


\begin{figure}[b]
    \includegraphics[width=\columnwidth]{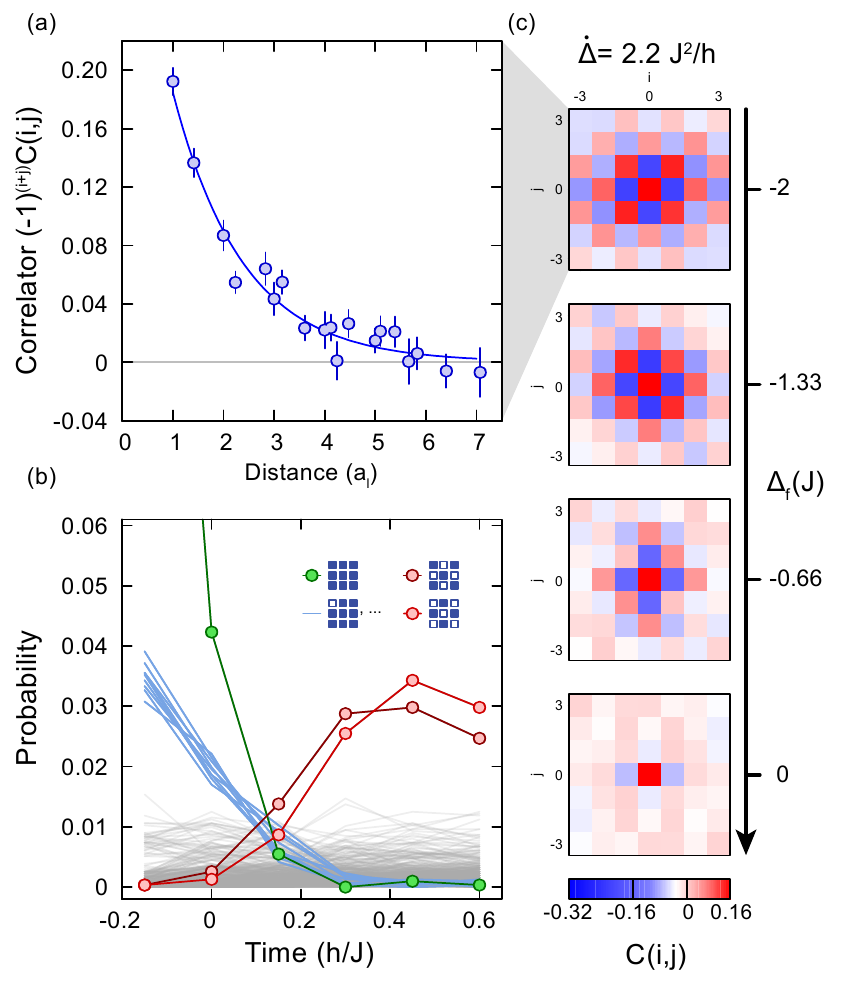}
    \caption{{\bfseries Characterizing many-body states during and after a slow quench.} (a) Spatial decay of the correlations after a sweep with $\dot{\Delta}=2.2J^2/h$, with an exponential fit that yields a correlation length $\xi=1.4(1)a_l$ sites. (b) Time evolution of the probabilities of observing different configurations in $3\times3$ sub-systems, not corrected for detection efficiencies. The probabilities are shown for the two antiferromagnetic states (red), the all-grounds state (green), one Rydberg atom states (blue), and all other states (grey). The evolution is shown during a ramp with $\dot{\Delta}=4.4J^2/h$. The antiferromagnetic configurations become most probable at the end of the quench. (c) Full correlation matrices $C(i,j)$ at different detunings during a quench from a paramagnet to antiferromagnet with $\dot{\Delta} = 2.2J^2/h$, showing the growing range of the antiferromagnetic correlations.\label{fig:fig4}}
\end{figure}

Near the end of the ramps, where significant antiferromagnetic correlations have built up, we find that we can fit the decay of the correlations with distance to an exponential (Fig.~\ref{fig:fig4}(a)), even though the spin system may not have reached thermal equilibrium. The fitted correlation lengths range from $\xi = 0.74(6)a_l$ to $\xi = 1.9(2)a_l$ depending on $\dot{\Delta}$. Another way to characterize the short range antiferromagnets created by these slow quenches is by extracting the probabilities for observing a particular spin configuration in a sub-system. In Fig.~\ref{fig:fig4}(b), we show the probability of observing different spin configurations in $3\times3$ sub-systems, not correcting for detection fidelities. The two antiferromagnetic states are the most probable states near the end of the ramp, with an enhancement of a factor of $16(2)$ over a uniform distribution in the Hilbert space. 


In conclusion, we studied quench dynamics in a 2D Ising model realized with ultracold atoms coupled to a Rydberg state in an optical lattice. The use of a light fermionic atom, $^6$Li, allows us to use Pauli blocking in a relatively large spacing lattice to create 2D atomic arrays with high-filling ($\sim96\%$), comparable to what is achieved in atom-by-atom assembler experiments \cite{Barredo2016,Endres2016}. Combining the large spacing with the use of a low-lying Rydberg state, we reached the strong correlation regime with $R_b\sim a_l$ and prepared states exhibiting strong short-range antiferromagnetic correlations. We found good agreement of our data with state-of-the-art numerics for short-time quench dynamics. Additionally, we studied the dynamics in a regime outside the reach of exact theoretical techniques, providing test data for approximate techniques to calculate dynamics. Our new ultracold $^6$Li Rydberg platform opens many interesting directions for future work. Rydberg excitation in a Fermi gas may allow the exploration of impurity dynamics in the presence of Pauli blocking effects \cite{Gaj2014,Schmidt2016}. Another possible direction is the use of Rydberg dressing techniques to realize a dipolar Fermi gas, taking advantage of the fast tunneling of lithium in an optical lattice to go beyond the frozen gas regime \cite{Xiong2014,Li2015,Jau2016,Zeiher2016}.

{\textit{Note:}} Recently, antiferromagnetic correlations have been observed in 2D arrays of Rydberg atoms trapped in optical tweezers in experiments at the Institut d'Optique.

\begin{acknowledgments}
This work was supported by the NSF (grant no. DMR-1607277), the David and Lucile Packard Foundation (grant no. 2016-65128), and the AFOSR Young Investigator Research Program (grant no. FA9550-16-1-0269). W.S.B. was supported by an Alfred P. Sloan Foundation fellowship. P.T.B. was supported by the DoD through the NDSEG Fellowship Program.
\end{acknowledgments}



%



\pagebreak
\clearpage
\setcounter{equation}{0}
\setcounter{figure}{0}

\renewcommand{\theparagraph}{\bf}
\renewcommand{\thefigure}{S\arabic{figure}}
\renewcommand{\theequation}{S\arabic{equation}}

\onecolumngrid
\flushbottom

\section{\large Supplemental Information}
\twocolumngrid

The quantum gas microscopy apparatus used for preparing the atomic arrays used in this work and site-resolved imaging of these arrays is described in detail in the supplements of ref.~\cite{Brown2017}.

\subsection{Laser system for coupling to Rydberg states}

For single-photon excitation to the Rydberg state, we use a deep UV laser system at 230 nm based on frequency-quadrupling light from a diode laser source (Laser \& Electro-Optic Solutions). We start with \SI{1.8}{\W} of \SI{920}{\nm} light from a tapered amplifier seeded by an external grating diode laser. Next, a frequency-doubling cavity generates \SI{750}{\mW} at \SI{460}{\nm}. Finally, a second doubling cavity yields \SI{\approx50}{\mW} of \SI{230}{\nm} light. A constant flow of oxygen was required to maintain intensity stability and increase the lifetime of the second harmonic generation crystal in the second cavity. The frequency of the \SI{920}{\nm} laser is stabilized to a Stable Laser Systems ultralow expansion glass (ULE) cavity with a measured finesse of $\approx4000$ and linewidth of \SI{370}{\kHz}. We use a fiber-EOM before the cavity to generate sidebands at two frequencies for Pound-Drever-Hall locking with offset frequency control of the laser with respect to the reference cavity \cite{Thorpe2008}. To obtain fast amplitude and frequency control of the UV light, we used an AOM (IntraAction ASM-1501LA61) driven by a fast VCO (MiniCircuits ZX95-200A+). 

\begin{figure}[h]
    \includegraphics[width=\columnwidth]{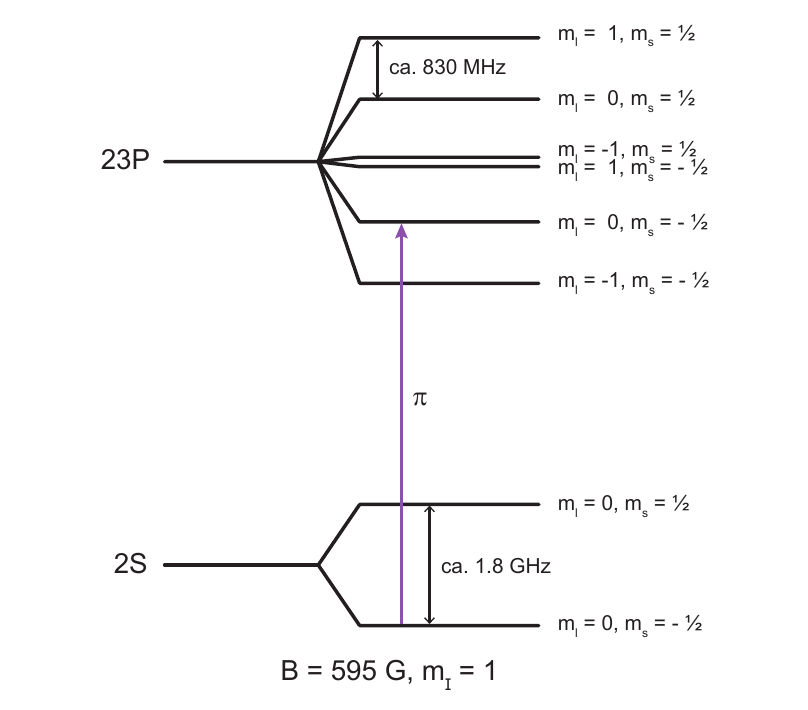}
    \caption{Level diagram showing laser coupling with linear ($\pi$) polarization from the ground state to the Rydberg state at a magnetic field of 595 G.}\label{fig:sfig2}
\end{figure}

For initial identification of the Rydberg lines, we performed a ``V-scheme" spectroscopy \cite{Thoumany2009} with the \SI{230}{\nm} beam and a diode laser driving the $D_2$ transition of lithium at \SI{671}{nm}. The locking point on the cavity was recorded for a range of Rydberg states including the $23P$ state used in this experiment (Fig.~\ref{fig:sfig2}). With the \SI{920}{\nm} laser locked, the final UV output had an intensity stability better than \SI{10}{\%}. The linewidth at \SI{230}{\nm} was measured to be \SI{\approx100}{\kHz}.

\subsection{Single atom Rabi oscillations and coherence}

\begin{figure}[h]
    \includegraphics[width=\columnwidth]{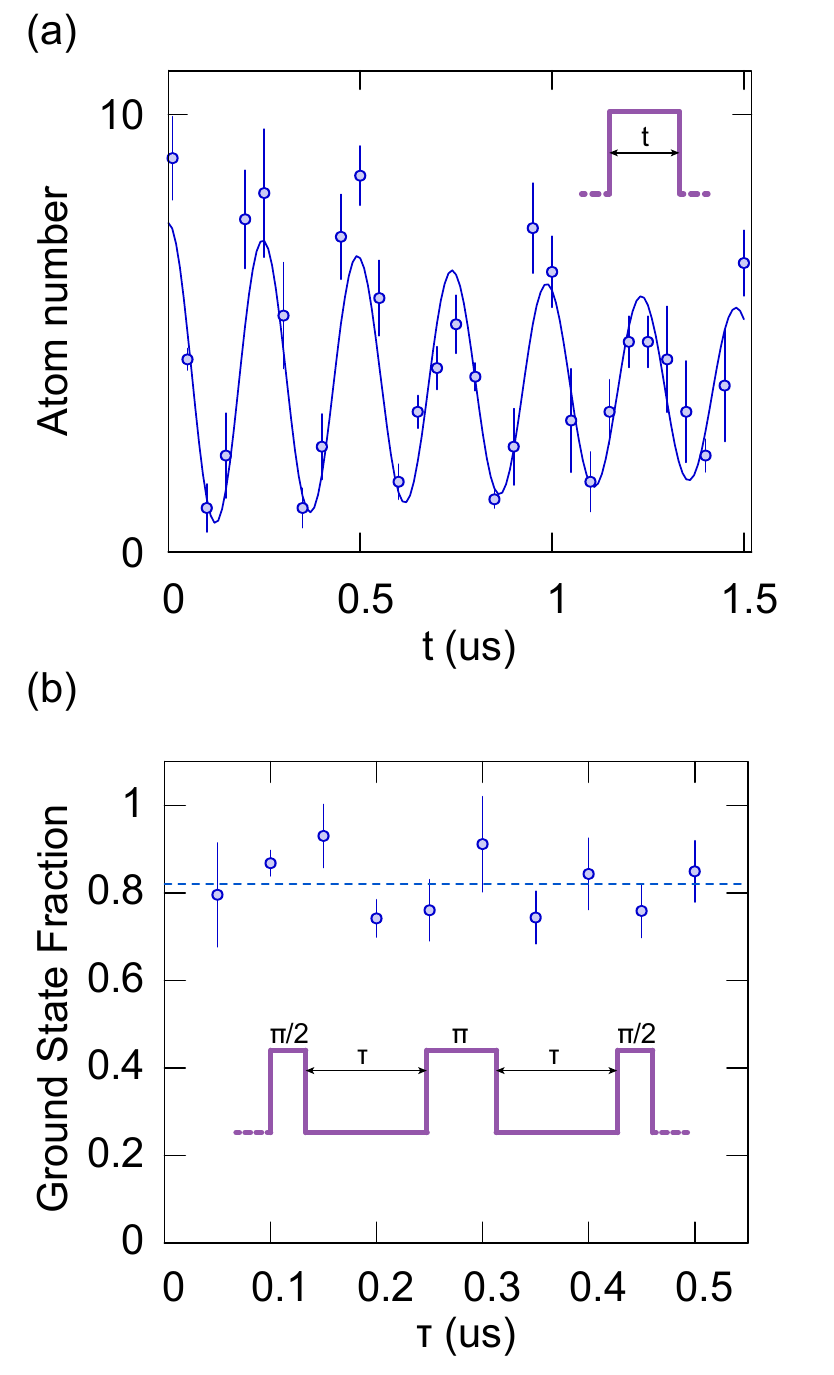}
    \caption{(a) Single atom Rabi oscillation. (b) Measurement of the ground state fraction after a Ramsey echo sequence, indicating no loss of coherence over times longer than the quench times used in this work.}\label{fig:sfig1}
\end{figure}

To measure the strength of the laser coupling to the Rydberg state, we measure Rabi oscillations in a sparse cloud where the interactions between the Rydberg atoms are negligible. A typical single atom Rabi oscillation is shown in Fig.~\ref{fig:sfig1}(a). The decay of the Rabi oscillation is mainly due to shot-to-shot fluctuations of the laser intensity. In addition we measure the coherence of the atoms in a sparse cloud using a Ramsey echo sequence: $\pi/2$ - $\tau$ - $\pi$ - $\tau$ - $\pi/2$ pulse, where $\tau$ is a delay time. The ground state fraction is measured at the end of the sequence (Fig.~\ref{fig:sfig1}(b)). Even for $\tau = 0$, corresponding to a 2$\pi$ pulse, the measured ground state fraction is reduced to $\sim0.8$, because of laser intensity fluctuations. However, for a total delay $2\tau=1~\mu$s, corresponding to $6h/J$, we do not observe any decay of the ground state fraction, indicating that there is no loss of coherence.

\subsection{Movement of the atoms during the quenches}

In our calculations, we ignore the movement of atoms during the quenches. In reality, the atoms can move because of strong van der Waals forces between atoms in Rydberg states, photon recoil kicks and anti-trapping forces on atoms in the Rydberg state. From the experimentally fitted value for $C_6$, we estimate Rydberg atoms accelerate towards each other with $a(r) = \frac{6 C_6}{r^7 m}$ where $a \approx \SI{3.2e6}{m/s^2}$ for neighboring Rydberg atoms and $a(\sqrt{2} a_\text{lat}) \approx \SI{2.8e5}{m/s^2}$ for next-neighbor Rydberg atoms. This leads to a displacement of 0.09$a_l$ (0.01$a_l$) for a typical quench time of 200 ns for nearest neighbor  (next-nearest neighbor) Rydberg atoms. We note that the excitation of two Rydberg atoms on neighboring sites is largely suppressed due to the blockade. Recoil and anti-trapping forces lead to much smaller displacements of the atoms.

\subsection{Theoretical calculation of $C_6$}

We calculate the value of $C_6$ for the state $23P,m_l=0,m_s=1/2$ at \SI{595}{G} with two different techniques. First we use a perturbation theory calculation in the $m_l$ basis which yields $C_6 = \SI{-1.915}{\MHz\,\um^6}$. As a second approach we use a pair state exact diagonalization in the $m_j$ basis at \SI{595}{G} using \cite{Weber2017} and perform a basis transformation to the $m_l$-basis. A fit to the potential curve then yields $C_6$. Depending on the inner cutoff $r_0$ for the fit we obtain values between $C_6 = \SI{-1.974}{\MHz\,\um^6}$ for $r_0=\SI{0.7}{\um}$ and $C_6 = \SI{-1.864}{\MHz\,\um^6}$ at $r_0 = \SI{1}{\um}$ which leads us to the error estimate in the main text.

\subsection{NLCE dynamics}
The numerical linked cluster expansion (NLCE) algorithm used to calculate the dynamics in this work is an extension of the NLCE technique for thermodynamic quantities, reviewed in refs~\cite{Rigol2007,Tang2013}. Our NLCE calculations take into account next-nearest neighbor (diagonal) interactions, lattice anisotropy, and the finite time for turning on and off the Rabi frequencies during ``sudden" quenches. We discuss some of the major modifications that are made to the algorithm, assuming the reader's familiarity with the standard algorithm.

For systems with the symmetry of the square lattice and only nearest neighbor interactions, each embedding of a graph on the lattice is dependent only on the topology of the graph, which allows for a significant reduction in the number of clusters that need to be diagonalized. Taking into account next-nearest neighbor interactions and lattice anisotropy means that this is no longer true, as two topologically identical graphs may have different graph Hamiltonians. Thus, we break down topologically identical graphs further into classes of graphs with the same Hamiltonian up to graph symmetries, each of which we only need to solve once.

For each of these, we then perform a time evolution starting from the initial state using sparse representations of the Hamiltonian. The initial and final ramp is simulated by breaking down the overall ramp time into five time steps, and applying the time evolution between these steps with the appropriate time-averaged Hamiltonian. Then, the appropriate correlation functions can be extracted from the final state. The subgraph subtraction then proceeds as usual, except that each embedding should be treated independently, as the contribution of a graph to a particular correlator depends on its embedding in the lattice. Finally, we perform an Euler resummation starting from the 3rd order to reduce odd-even order fluctuation. Note that the graphical expansion done here is site-based, rather than link-based.

Finally, we have checked the effect of including beyond-next-nearest neighbor interactions, and the number of time steps, and found that our results are well converged with respect to them. 

\subsection{Fitting sudden quench dynamics to NLCE results}

The correlator dynamics are computed using NLCE for a grid of $\Delta$ and $C_6$ values at 9th order in the expansion, taking into account the independently calibrated Rabi frequency $\Omega = h\times\SI{4.05(2)}{\MHz}$. The correlators $C(0,0)$, $C(1,0)$, and $C(0,1)$ are simultaneously fit to the results using two fit parameters: $C_6$ and a scaling factor $\alpha$ corresponding to a Rydberg atom detection efficiency.  The scaling factor reduces the nearest neighbor correlators $C(1,0)$ and $C(0,1)$ as $\alpha^2$. Since $C(0,0) = \avg{n^2} - \avg{n}^2$ and $n^2 = n$ since $n$ is either 0 or 1, we obtain $C(0,0) = \avg{n} - \avg{n}^2$. This leads to a corrected correlator $C^*(0,0) = \alpha\avg{n} - \alpha^2\avg{n}^2$.

\end{document}